\documentstyle{article}
\begin{document}

\def\thefootnote{\fnsymbol{footnote}}

\def\diam{{\hbox{\hskip-0.02in
\raise-0.090in\hbox{$\displaystyle\bigvee$}\hskip-0.200in
\raise0.099in\hbox{ $\displaystyle{\bigwedge}$}}}}
\def\dop{\mathop{{\diam}}\limits}
\def\bw#1#2#3#4{
{\scriptstyle{#4}}\,
{\dop_{#3}^{#1}}
{\scriptstyle{#2}}  }
\overfullrule=0pt


\def\min{\rm min}
\def\q#1{\left[ #1 \right]}
\def\s#1#2{S_{#1,#2}}
\def\b#1#2#3#4{B_{#1,#2} \left[{{\phi_{1} \phi_{1} }\atop {\phi_{#3} \phi_{#4}}}\right] }
\def\B#1#2#3#4#5#6{B_{#1,#2} \left[{{\phi_{#5} \phi_{#6} }\atop {\phi_{#3} \phi_{#4}}}\right] }
\def\C#1#2#3#4#5#6{C_{#1,#2} \left[{{\phi_{#5} \phi_{#6} }\atop {\phi_{#3} \phi_{#4}}}\right] }
\def\conf#1{\Delta_{#1}}
\def\P#1#2#3{{\bf P}^{(#1)}_{#2,#3}}
\def\bb#1#2#3#4{B_{\phi_{#1},\phi_{#2}} \left[{{\phi_{1} \phi_{1} }\atop {\phi_{#3} \phi_{#4}}}\right] }
\def\BB#1#2#3#4#5#6{B_{\phi_{#1},\phi_{#2}} \left[{{\phi_{#5} \phi_{#6} }\atop {\phi_{#3} \phi_{#4}}}\right] }
\def\cor#1#2#3#4{\langle \phi_{#1} \phi_{#2} \phi_{#3} \phi_{#4} \rangle}
\def\w#1#2#3#4#5{w \left( {{{#1 \hskip 0.2cm #2} \atop {#3 \hskip 0.2cm #4}} \bigg\vert } {#5} \right)}
\def\wl#1#2#3#4#5{{\overline w} \left( {{{#1 \hskip 0.2cm #2} \atop {#3 \hskip 0.2cm #4}} \bigg\vert } {#5} \right)}

\def\el#1#2#3#4{{[ #1 ] [ #2 ] } \over {[ #3 ] [ #4 ]}}

\begin{center}

\hfill    WIS-96/28/July-PH\\
\hfill       hep-th/

\vskip 1 cm

{\large \bf  New Solvable Lattice Models from Conformal Field Theory}

\vskip 1 cm

Ernest Baver 

\vskip 1 cm

{\em Department of Particle Physics\\
The Weizmann Institute\\
Rehovot 76100\\
ISRAEL}

\end{center}

\vskip 1 cm

\begin{abstract}

In this work we build the trigonometric solutions of the Yang-Baxter equation that can not be obtained from quantum groups in any direct way. The solution is obtained using the construction suggested recently \cite{gep} from the rational conformal field theory corresponding to the WZW model on $SO(3)_{4 R}=SU(2)_{4 R} / Z_{2}$.
We also discuss the full elliptic solution to the Yang-Baxter equation whose critical limit corresponds to the trigonometric solution found below.

\end{abstract}



\section{Introduction}

Two dimensional systems offer the intriguing possibility of exact solvabillity. The solutions to the Yang-Baxter equation play a central role in theory of the integrable systems \cite{general}. Quantum groups yield the  systematic approach of the construction of the trigonometric solutions of the Yang-Baxter equation  \cite{delius}. In this work we study the trigonometric solutions of the Yang-Baxter equation that can not be obtained from quantum groups in any direct way. We also discuss the full elliptic solution to the Yang-Baxter equation whose critical limit corresponds to the trigonometric solution found below.

We use the construction suggested recently \cite{gep} stating that solvable lattice models may be build around any rational conformal field theory together with some primary field $\phi$. In particular in this way the models found previously \cite{jimbo} were rederived from conformal field theory \cite{gepfuchs,gep1} in a systematic way.

We will deal here with the interaction round the face (IRF) lattice models. 

The partition function of the lattice model is given by

$$Z=\sum_{{\small \rm configurations}} \prod_{{\small \rm faces}} \w{a}{b}{c}{d}{u}   . \eqno(1.1)$$The model is solvable if the Boltzmann weights (BW) $\w{a}{b}{c}{d}{u}$ obey the Yang-Baxter equation:

{\small $$\sum_{c} \w{b}{d}{a}{c}{u} \w{a}{c}{g}{f}{u+v} \w{c}{d}{f}{e}{v}=\sum_{c} \w{a}{b}{g}{c}{v} \w{b}{d}{c}{e}{u+v} \w{c}{e}{g}{f}{u}. \eqno(1.2)$$}The construction described in \cite{gep} states that BW are given in terms of braiding matrices of the corresponding RCFT as:

$$\w{a}{b}{c}{d}{u} \equiv \bw{a}{b}{d}{c}=\sum_{j} \langle a,b,d | P^{(j)} | a,c,d \rangle \rho_{j} (u), \eqno(1.3)$$

$$\langle a,b,d | P^{(j)} | a,c,d \rangle = \prod_{j \not = l} {{\B{b}{c}{a}{d}{}{}-\delta_{b,c} \lambda_{j}} \over {\lambda_{l}-\lambda_{j}} },\eqno(1.4)$$Here $j$ labels the field exchanged in the u-channel. The fields that may appear in the u-chanell are determined from the product of the field $\phi$ with itself $\phi \times \phi=\sum_{j} N_{\phi,\phi}^{j} \phi_{j}$, where $N_{j,k}^{l}$ are the fusion coefficients of the RCFT under consideration and $\rho_{j}(u)$ are some scalar functions which will be specified below. Projection operators $P^{(j)}$ are found from the braiding matrix $\B{b}{c}{a}{d}{}{}$ whose eigenvalues $\lambda_{j}$ are given by $\lambda_{j}=(-1)^{j} e^{i \pi (\Delta_{j}-2 \Delta_{\phi})}$, where $\Delta,\Delta_{j}$  are conformal dimensions of the primary fields $\phi,\phi_{j}$. Note that the Boltzmann weights vanish, unless the  admissibility condition is satisfied:

$$N_{a,\phi}^{b} N_{b,\phi}^{c} N_{c,\phi}^{d} N_{d,\phi}^{a} >0. \eqno(1.5) $$

In this paper we will derive the solvable lattice model based on the extended current algebra of WZW model on $SO(3)_{4 R}=SU(2)_{k=4 R}/Z_2$ with the primary field $\phi_{1}$ (adjoint). Its fusion rules and torus modular $S$ matrix were derived in \cite{extfus,bg}. We quote the result here for the torus modular $S$ matrix for $R$ even

$$S={\pmatrix{2 \s{j}{i}&\s{j}{R} &\s{j}{R^{\prime}} \cr \s{R}{i} &
x&z \cr \s{R^{\prime}}{i} &z&x}}, \hskip 0.8cm x={{S_{R,R}+1} \over 2}, \hskip 0.8cm z={{S_{R,R}-1} \over 2}, \eqno(1.6)$$where $S_{a,b}={\sqrt{1 \over {2 R+1}}} \sin{{{(2 a+1)(2 b+1) } \over {4 R+2}} \pi}$. The entries of this matrix are restricted to the singlets of $Z_2$ (only integer isospins may appear) and due to the symmetry $\s{j}{i}=\s{\sigma(j)}{i}=\s{j}{\sigma(i)}$, where $\sigma$ is the external automorphism $\sigma(j)=2 R-j$,  the primary fields are labeled modulo\footnote{For the sake of concreteness we chose to label primary fields by minimal isospin $\min(j, 2 R-j)$, so that primary fields of the theory are $\phi_{0}$ (identity primary field), $\phi_{1},...,\phi_{R-1},\phi_{R},\phi_{R^{\prime}}$} $\sigma$. The fixed point of $\sigma$ is resolved into two fields: $R,R^{\prime}$, these fields differ by an additional quantum number but have the same conformal weights and transformation properties under $SU(2)$. The fusion rules for $R$ even are given by:

$$\phi_{j} \times \phi_{i} =\sum_{p=|i-j|}^{\min (R,i+j)}m_{p} \phi_{p}, \hskip 0.8cm i,j {\not=} R,R^{\prime}, \eqno(1.7)$$where $m_{R}=m_{R^{\prime}}=1$ and $m_{p}=2$  iff $p,2 R-p \in \{|i-j|,...,i+j\}$.

$$\phi_{R} \times \phi_{R}=\phi_{R^{\prime}} \times \phi_{R^{\prime}}=\sum_{p=0}^{R \over 2} \phi_{2 p}, \eqno(1.8)$$

$$\phi_{R} \times \phi_{R^{\prime}} =\sum_{p=0}^{{R-2} \over 2} \phi_{2 p+1}, \eqno(1.9)$$

$$\phi_{j} \times \phi_{R} =\sum_{p=|j-R|}^{R-1} \phi_{p}+{{1+{(-1)}^{j}} \over 2 } \phi_{R}+{{1-{(-1)}^{j}} \over 2 } \phi_{R^{\prime}}. \hskip 0.8cm j {\not=} R,R^{\prime} \eqno(1.10)$$The scalar functions $\rho^{(j)} (u)$ in this case are given by \cite{gep}:

$$\rho_{0} (u)={{\sin(\lambda-u) \sin(\omega-u)} \over {\sin(\lambda) \sin(\omega)}}, \hskip 0.8cm \rho_{1} (u)={{\sin(\lambda+u) \sin(\omega-u)} \over {\sin(\lambda) \sin(\omega)}}, \eqno(1.11)$$

$$\rho_{2} (u)={{\sin(\lambda+u) \sin(\omega+u)} \over {\sin(\lambda) \sin(\omega)}},   \eqno(1.12) $$where $\lambda,\omega$ are the crossing parameters of the model that are related to the conformal weights of the fields appearing in the operator product expansion of field $\phi_{1}$ with itself: $\phi_{1} \times \phi_{1}={\bf 1}+\phi_{1}+\phi_{2}$

$$\lambda={\pi \over 2} (\Delta_{1}-\Delta_{0})={{\pi} \over {4 R+2}}, \hskip 0.8cm \omega={\pi \over 2} (\Delta_{2}-\Delta_{1})=2 \lambda. \eqno(1.13)$$

The incidence diagram of matrix $N_{a,\phi_{1}}^{b}$ \cite{bg} representing the  admissibility condition for the lattice variables is shown at Fig.1

\begin{figure}[ht]
\unitlength=1.00mm
\linethickness{0.8pt}

\begin{picture}(129.60,92.00)

\put(33.00,80.00){\line(1,0){51.00}}
\put(85.00,80.00){\line(2,1){15.00}}
\put(85.00,80.00){\line(2,-1){15.00}}
\put(100.00,72.00){\line(0,1){16.00}}

\put(85.00,80.00){\circle*{2}}
\put(72.00,80.00){\circle*{2}}
\put(59.00,80.00){\circle*{2}}
\put(46.00,80.00){\circle*{2}}
\put(33.00,80.00){\circle*{2}}
\put(100.00,87.00){\circle*{2}}
\put(100.00,73.00){\circle*{2}}

\put(85.00,84.00){\oval(4,4)[t]}
\put(83.00,84.30){\line(2,-5){2.00}}
\put(84.30,79.30){\line(3,5){3.00}}

\put(72.00,84.00){\oval(4,4)[t]}
\put(70.00,84.30){\line(2,-5){2.00}}
\put(71.30,79.30){\line(3,5){3.00}}

\put(59.00,84.00){\oval(4,4)[t]}
\put(57.00,84.30){\line(2,-5){2.00}}
\put(58.30,79.30){\line(3,5){3.00}}

\put(46.00,84.00){\oval(4,4)[t]}
\put(44.00,84.30){\line(2,-5){2.00}}
\put(45.30,79.30){\line(3,5){3.00}}

\put(105.00,90.00){\makebox(0,0)[cb]{{\small ${\bf R}$}}}
\put(105.00,70.00){\makebox(0,0)[cb]{{\small${\bf R^{\prime}}$}}}
\put(85.00,74.00){\makebox(0,0)[cb]{{\small${\bf R-1}$}}}
\put(72.00,74.00){\makebox(0,0)[cb]{\small ${\bf R-2}$}}

\put(46.00,74.00){\makebox(0,0)[cb]{\small ${\bf 1}$}}
\put(33.00,74.00){\makebox(0,0)[cb]{\small ${\bf 0}$}}

\put(63.00,60.00){\makebox(0,0)[cb]{\small ${\rm Fig.1 \hskip 0.5cm Admisibility \hskip 0.5cm graph}$}}

\put(63.00,50.00){\makebox(0,0)[cb]{\small  Here $R$  and  $R^{\prime}$ are two primary fields corresponding to the fixed point \cite{bg}  }}

\end{picture}
\end{figure}


\section{Trigonometric Solution}
\subsection{Boltzmann Weights without Fixed Point Fields} 

We will start by exploiting the connection between our model and the model described at \cite{pasq} which in this language corresponds to the IRF model build around $SU(2)_{k=4 R}$ together with the field $\phi_{1}$. We will refer in the sequel to this model restricted to the singlets of $Z_{2}$ as the diagonal model.

Let us denote by $F_{p}^{ijkl} (f_{p}^{ijkl})$ correspondingly the conformal blocks of extended (unextended) theory, the corresponding braiding matrices are denoted by $B$ and $C$. Below we assume for the moment that none of the fields is equal to the fixed point field $\phi_{R}$

$$F_{p}^{ijkl}=\sum_{\sigma} f_{p}^{\sigma(ijkl)}, \eqno(2.1)$$
$$F_{p}^{ijkl}=\sum_{p^{\prime}} \B{p}{p^{\prime}}{i}{l}{j}{k} F_{p^{\prime}}^{ijkl}, \eqno(2.2)$$ 

$$f_{p}^{ijkl}=\sum_{p^{\prime}} \C{p}{p^{\prime}}{i}{l}{j}{k} f_{p^{\prime}}^{ijkl}, \eqno(2.3)$$so using only definitions

$$\sum_{p^{\prime}} \B{p}{p^{\prime}}{i}{l}{j}{k} F_{p^{\prime}}^{ijkl}=\sum_{\sigma} \sum_{p^{\prime}} C_{p,p^{\prime}}{ \left[ \sigma \left( {{j k} \atop {i l}} \right) \right]} f_{p^{\prime}}^{\sigma (ijkl) }, \eqno(2.4)$$where in the previous equation by $\sigma (i,j,k,l)$ we mean some collection of Dynkin labels $\sigma (i), \sigma^{\prime} (j), \sigma^{\prime \prime} (k), \sigma^{\prime \prime \prime} (l)$ and $\sigma \in Z_{2}$. Of course some of the conformal blocks $f_{p^{\prime}}^{\sigma (ijkl) }$ in the RHS of Eq.(2.4) may vanish. Note that Eqs.(2.1-2.4) imply the relation of the form $\sum_{j} b_{j} f_{j}(z)  =\sum_{j} c_{j} f_{j} (z)$, where $f_{j}$ are independent functions and $c_{j},b_{j}$ are some coefficients. From this follows the equality of the coefficients, namely we have

$$\B{p}{p^{\prime}}{i}{l}{j}{k}=C_{p,p^{\prime}}{ \left[{{j k} \atop {i l}} \right]}, \hskip 0.8cm p,p^{\prime},i,j,k,l \not = R.  \eqno(2.5)$$It means that the Boltzmann weights which do not contain fixed points fields are equal to the corresponding Boltzmann weights of the model described at \cite{pasq}.

\subsection{Boltzmann Weights Involving Fixed Points}

First we calculate the braiding matrix $\b{p}{q}{R}{R^{\prime}}$. Note that corelator $\cor{R}{1}{1}{R^{\prime}}$ receives contribution in the s-channel only from the field $\phi_{R-1}$, so that corresponding braiding matrix $\b{p}{q}{R}{R^{\prime}}$ is fixed by monodromy invariance $\b{p}{q}{R}{R^{\prime}}=e^{2 \pi i (\Delta_{R}-\Delta_{R-1})}$.

Now we  calculate  the braiding matrix $\B{p}{q}{R}{R}{1}{1}$. From the fusion rules we know that the relevant space of conformal blocks is two dimensional: $p,q=\phi_{R-1},\phi_{R^{\prime}}$, so that it is enough to find only one entry of the braiding matrix and the others will be fixed from the invariance under monodromy. Using pentagon identity \cite{ms} \footnote{This form of pentagon identity is obtained from the conventional using compatibility between braiding and fusing.}

{\small {$$\B{p_1}{q_1}{i_1}{j_3}{j_2}{p_2} \B{p_2}{q_2}{j_1}{j_5}{q_1}{j_4} e^{i \pi (\conf{p_1}-\conf{q_2}-\conf{j_1})}=$$
$$=\sum_{s} e^{i \pi \conf{s}} \B{p_2}{s}{p_1}{j_5}{j_3}{j_4} \B{p_1}{q_2}{j_1}{s}{j_2}{j_4} \B{s}{q_1}{q_2}{j_3}{j_2}{j_5}, \eqno(2.6)$$}} 

For R-even let us set 

$$j_{2}=p_{2}=j_{4}=j_{5}=\phi_{1}, \eqno(2.7)$$
  $$j_{1}=j_{3}=\phi_{R}, \hskip 0.5cm p_{1}=q_{1}=\phi_{R}^{\prime}, \hskip 0.5cm q_{2}=\phi_{R-1}. \eqno(2.8)$$So that from pentagon identity we have, because in RHS only $s=\phi_{R-1}$ will survive

$$\b{\phi_{R}^{\prime}}{\phi_{R}^{\prime}}{R}{R}=\b{\phi_{R}^{\prime}}{\phi_{R-1}}{R}{R-1} \b{\phi_{R-1}}{\phi_{R}^{\prime}}{R-1}{R}, \eqno(2.9)$$For the matrices in the RHS we may use the expression which was found for arbitrary $2 \times 2$ braiding matrix\footnote{We did not use this expression directly, because exponent sum rule in this case is not obeyed \cite{gepfuchs}} \cite{gepfuchs}. Other entries are calculated from the monodromy invariance

$$\sum_{p^{\prime}} e^{2 \pi i (\Delta_{p}+\Delta_{p^{\prime}})} \b{p}{p^{\prime}}{R}{R} \b{p^{\prime}}{q}{R}{R}=\delta_{p,q} e^{4 \pi i \Delta_{R}}, \eqno(2.10)$$After some algebra we have:

$$\b{p}{q}{R}{R}={1 \over \q{2 R}^2} \pmatrix{{1 \over q} \q{2R+2} \q{2R-2} & * \cr {q^{R}} \sqrt {1-{{\q{2R+2}^2 \q{2R-2}^2} \over \q{2R}^4}} & \q{2R+2} \q{2R-2}}, \eqno(2.11)$$where $\q{z} \equiv {{q^{z \over 2}-q^{-{z \over 2}}} \over {q^{1 \over 2}-q^{-{1 \over 2}}}}, \hskip 0.3cm q=e^{\pi i \over {2 R+1}}$.

Now we will turn to the calculation of the braiding matrix $\b{p}{q}{R-1}{R-1}$, it is $4 \times 4$ matrix (the relevant space of conformal blocks is 4 dimensional $p,q=\phi_{R-2}, \phi_{R-1}, \phi_{R}, \phi_{R^{\prime}}$). Using twice pentagon identity we have:

{\small $$\bb{R}{R}{R-1}{R-1} \BB{1}{R^{\prime}}{R-1}{1}{R}{1} e^{2 \pi i (\Delta_{R}-\Delta_{R-1})}=$$

$$=\BB{1}{R-1}{R}{1}{R-1}{1} \bb{R}{R^{\prime}}{R-1}{R-1} \bb{R-1}{R}{R^{\prime}}{R-1}, \eqno(2.12)$$}and similarly

{\small $$\bb{R}{R^{\prime}}{R-1}{R-1} \BB{1}{R}{R-1}{1}{R^{\prime}}{1} e^{2 \pi i (\Delta_{R}-\Delta_{R-1})}=$$

$$=\BB{1}{R-1}{R}{1}{R-1}{1} \bb{R}{R}{R-1}{R-1} \bb{R-1}{R^{\prime}}{R}{R-1}. \eqno(2.13)$$}From these equations one may show: 

$$\b{R}{R}{R-1}{R-1}= \pm \b{R}{R^{\prime}}{R-1}{R-1}. \eqno(2.14)$$Now using monodromy invariance 

$$\sum_{p^{\prime}} e^{2 \pi i (\Delta_{p}+\Delta_{p^{\prime}})} \b{p}{p^{\prime}}{R-1}{R-1} \b{p^{\prime}}{q}{R-1}{R-1}=\delta_{p,q} e^{4 \pi i \Delta_{R-1}}, \eqno(2.15)$$we find

$$\b{R}{R}{R-1}{R-1}=-\b{R}{R^{\prime}}{R-1}{R-1}={1 \over 2} e^{2 \pi i (\Delta_{R-1}-\Delta_{R})}, \eqno(2.16)$$Using Eqs. (1.3-1.4, 1.11-1.13) one may find the Boltzmann weights of the model. Let us consider for example the $\bw{R}{R-1}{R^{\prime}}{R-1}$, the only  field exchanged in the u-channel is $\phi_{1}$ so that the corresponding BW is given by:

$$\bw{R}{R-1}{R^{\prime}}{R-1}=\rho_{1}(u), \eqno(2.17)$$after some straitforward calculation the rest of the Boltzmann Weights are found to be: 

$$\bw{R}{R-1}{R}{R-1}=\bw{R^{\prime}}{R-1}{R^{\prime}}{R-1}=2 \rho_{0}(u) \P{0}{R-1}{R-1} +2 \rho_{2}(u) \P{2}{R-1}{R-1}, \eqno(2.18)$$

$$\bw{R}{R^{\prime}}{R}{R-1}=\bw{R}{R-1}{R}{R^{\prime}}={\sqrt 2} \rho_{0}(u) \P{0}{R}{R-1}  +{\sqrt 2} \rho_{2}(u) \P{2}{R}{R-1}, \eqno(2.19) $$

$$\bw{R-1}{b}{R-1}{R,R^{\prime}}={\rho_{0} (u) \over \sqrt 2} \P{0}{b}{R} + {\rho_{1} (u) \over \sqrt 2} \P{1}{b}{R} + {\rho_{2} (u) \over \sqrt 2} \P{2}{b}{R} , \hskip 0.8cm b \not =R,R^{\prime}, \eqno(2.20) $$

$$\bw{R-1}{R}{R-1}{R}={\rho_{0} (u) \over 2} \P{0}{R}{R} + {\rho_{1} (u) \over 2} \P{1}{R}{R} + {\rho_{2} (u) \over 2} (\P{2}{R}{R}+1) , \eqno(2.21) $$

$$\bw{R-1}{R}{R-1}{R^{\prime}}={\rho_{0} (u) \over 2} \P{0}{R}{R} + {\rho_{1} (u) \over 2} \P{1}{R}{R} + {\rho_{2} (u) \over 2} (\P{2}{R}{R}-1), \eqno(2.22) $$

$$\bw{R}{R}{R}{R}=\bw{R^{\prime}}{R^{\prime}}{R^{\prime}}{R^{\prime}}=0, \eqno(2.23)$$
$$\bw{R^{\prime}}{R}{R^{\prime}}{R}=\bw{R}{R^{\prime}}{R}{R^{\prime}}=\rho_{0} (u) \P{0}{R}{R}+\rho_{1} (u) \P{1}{R}{R}+\rho_{2} (u) \P{2}{R}{R} , \eqno(2.24)$$where $\P{j}{b}{c} \equiv \langle a,b,d | P^{(j)} | a,c,d \rangle$ are projectors from \cite{pasq} (their explicit expression are summarized in the appendix) and scalar functions $\rho_{j} (u)$ are given by:

$$\rho_{0} (u)={{\sin(\lambda-u) \sin(\omega-u)} \over {\sin(\lambda) \sin(\omega)}}, \hskip 0.8cm \rho_{1} (u)={{\sin(\lambda+u) \sin(\omega-u)} \over {\sin(\lambda) \sin(\omega)}}, \eqno(2.25)$$

$$\rho_{2} (u)={{\sin(\lambda+u) \sin(\omega+u)} \over {\sin(\lambda) \sin(\omega)}}, \hskip 0.8cm \lambda={1 \over 2} \omega={{\pi} \over {4 R+2}}.  \eqno(2.26) $$

Note that the Boltzmann weights obey the following crossing relation \cite{gep}:

$$\w{a}{b}{c}{d}{u}=\sqrt{{S_{b,0} S_{d,0}} \over {S_{a,0} S_{c,0}}} \w{a}{b}{c}{d}{-\lambda-u}, \eqno(2.27)$$with the torus modular matrix $S$ for the extended theory Eq.(1.6), where ``0" designates the identity primary field.

\section{Thermalized Boltzmann Weights}

In this section we will present the full off critical solution. The Boltzmann weights are now parameterized by the elliptic theta functions $\Theta_{1} (u,p)$ which is defined by:

$$\Theta_{1} (u,p)=2 p^{1 \over 4} \sin{u} \prod_{n=1}^{\infty} (1-2 p^{2 n} \cos(2 u) +p^{4 n}) (1-p^{2 n}) \equiv [u], \eqno(3.1)$$where p labels the distance from criticality. In the limit $p \rightarrow 0$ the trigonometric solution of the previous section is recovered.

The related diagonal model was found at \cite{jimbo} \footnote{We use here the fact that the model corresponding to the vector representation of $B_{1}$ is equivalent to the symmetric tensor of degree 2 $A_{1}$ model} The Boltzmann weights that do not contain fixed point fields are remained unchanged:

$$\bw{j}{j+1}{j+2}{j+1}={\el{\lambda+u}{\omega+u}{\lambda}{\omega}}, \eqno(3.2)$$

$$\bw{j}{j+1}{j+1}{j+1}={\el{\lambda+u}{(j+1) \omega-u}{\lambda}{(j+1) \omega}}, \eqno(3.3)$$

$$\bw{j}{j}{j+1}{j}={\el{\lambda+u}{j \omega+u}{\lambda}{j \omega}}, \eqno(3.4)$$

$$\bw{j}{j+1}{j+1}{j}={{[\lambda+u][u] \over [\lambda][\omega]}} {{\sqrt{[(j+2)\omega][j \omega]} \over [(j+1) \omega]}}. \eqno(3.5)$$

$$\bw{j}{j+1}{j}{j-1}={\el{u}{\lambda+u -\omega}{\lambda}{\omega}} {{\sqrt{[(j+{3 \over 2}) \omega][(j-{3 \over 2}) \omega]}} \over [(j+{1 \over 2}) \omega]}, \eqno(3.6)$$

{\small $$ \bw{j}{j+1}{j}{j+1}= {\el{\lambda-u}{(2 j+1) \omega-u}{\lambda}{(2 j+1) \omega}}+{\el{u}{(2 j+{3 \over 2}) \omega-u)}{\lambda}{(2 j+1) \omega}} {[j \omega] \over [(j+1) \omega]}, \eqno(3.7)$$}

{\small $$ \bw{j}{j-1}{j}{j-1}= {\el{\lambda+u}{2 j \omega+u}{\lambda}{2 j \omega}}-{\el{u}{2 j \omega+\lambda+u)}{\lambda}{2 j  \omega}} {[(j-{1 \over 2}) \omega] \over [(j+{1 \over 2}) \omega]}, \eqno(3.8)$$}

{\small $$\bw{j}{j}{j}{j}={\el{\lambda-u}{(j+{1 \over 2}) \omega-u}{\lambda}{(j+{1 \over 2}) \omega}}+{\el{u}{(j+1) \omega-u}{\lambda}{(j+{1 \over 2}) \omega}} ({\el{j \omega}{(j+{3 \over 2}) \omega}{(j+1) \omega}{(j+{1 \over 2}) \omega}}+{\el{(j+1) \omega}{(j-{1 \over 2}) \omega}{j \omega }{(j+{1 \over 2}) \omega}}), \eqno(3.9)$$}

The Boltzmann weights containing fixed point fields are given by:

{\small $$ \bw{R-1}{R^{\prime}}{R-1}{R^{\prime}}={1 \over 2} {\el{\lambda-u}{(2 R-1) \omega-u}{\lambda}{(2 R-1) \omega}}+{1 \over 2} {\el{u}{(2 R-{1 \over 2}) \omega-u)}{\lambda}{(2 R-1) \omega}} {[(R-1) \omega] \over [R \omega]}+{1 \over 2} {\el{\lambda+u}{\omega+u}{\lambda}{\omega}}, \eqno(3.10)$$

$$\bw{R-1}{R^{\prime}}{R-1}{R}={1 \over 2} {\el{\lambda-u}{(2 R-1) \omega-u}{\lambda}{(2 R-1) \omega}}+{1 \over 2} {\el{u}{(2 R-{1 \over 2}) \omega-u)}{\lambda}{(2 R-1) \omega}} {[(R-1) \omega] \over [R \omega]}-{1 \over 2} {\el{\lambda+u}{\omega+u}{\lambda}{\omega}}, \eqno(3.11)$$}

$$\bw{R}{R-1}{R^{\prime}}{R-1}={\el{\lambda+u}{\omega-u}{\lambda}{\omega}}, \eqno(3.12)$$

{\small $$\bw{R^{\prime}}{R}{R^{\prime}}{R}={\el{\lambda-u}{(R+{1 \over 2}) \omega-u}{\lambda}{(R+{1 \over 2}) \omega}}+{\el{u}{(R+1) \omega-u}{\lambda}{(R+{1 \over 2}) \omega}} ({\el{R \omega}{(R+{3 \over 2}) \omega}{(R+1) \omega}{(R+{1 \over 2}) \omega}}+{\el{(R+1) \omega}{(R-{1 \over 2}) \omega}{R \omega }{(R+{1 \over 2}) \omega}}). \eqno(3.13)$$}

The Yang-Baxter equation may be proved easily if one notes the following relations between the Boltzmann Weights of our model and the corresponding model of \cite{jimbo}. Indeed we have:

$$\bw{R}{R-1}{R^{\prime}}{R-1}=\bw{R}{R-1}{R}{R-1}-\bw{R}{R-1}{R}{R+1}, \eqno(3.14)$$
$$\bw{R}{R-1}{R}{R-1}=\bw{R}{R-1}{R}{R-1}+\bw{R}{R-1}{R}{R+1}, \eqno(3.15)$$

$$\bw{R-1}{R^{\prime}}{R-1}{R}={1 \over 2}(\bw{R-1}{R}{R-1}{R}-\bw{R-1}{R}{R+1}{R}), \eqno(3.16)$$
$$\bw{R-1}{R^{\prime}}{R-1}{R^{\prime}}={1 \over 2} (\bw{R-1}{R}{R-1}{R}+\bw{R-1}{R}{R+1}{R}), \eqno(3.17)$$

$$\bw{R-1}{j}{R-1}{R^{\prime}}={1 \over \sqrt{2}} \bw{R-1}{j}{R-1}{R} , \hskip 0.8cm j {\not =} R,R^{\prime} \eqno(3.18)$$
$$\bw{R}{R-1}{R}{R^{\prime}}=\sqrt{2} \bw{R}{R-1}{R}{R}, \eqno(3.19)$$

$$\bw{R}{R^{\prime}}{R}{R^{\prime}}=\bw{R}{R}{R}{R}, \eqno(3.20)$$where the Boltzmann Weights in the RHS of the Eqs.(3.14-3.20) are that of the corresponding diagonal model.

Let us consider as an example the following equation:

{\small $$\sum_{p} \w{R}{p}{R-2}{R-1}{u} \w{R}{R-1}{R^{\prime}}{p}{v} \w{p}{R^{\prime}}{R-1}{R-2}{u+v}=$$
$$=\sum_{p} \w{R}{R-1}{p}{R-1}{u+v} \w{R-1}{R^{\prime}}{R-1}{p}{u} \w{R-1}{p}{R-1}{R-2}{v},\eqno(3.21)$$}Using Eqs.(3.14-3.20) one may show that this equation is equivalent up to a factor ${1 \over \sqrt 2}$ to the difference of the two following Yang-Baxter equations which should hold as was proved in \cite{jimbo}:

 {\small $$\sum_{p} \w{R}{p}{R-2}{R-1}{u} \w{R}{R-1}{R}{p}{v} \w{p}{R}{R-1}{R-2}{u+v}=$$
$$=\sum_{p} \w{R}{R-1}{p}{R-1}{u+v} \w{R-1}{R}{R-1}{p}{u} \w{R-1}{p}{R-1}{R-2}{v},\eqno(3.22)$$}

{\small $$\sum_{p} \w{R}{p}{R-2}{R-1}{u} \w{R}{R+1}{R}{p}{v} \w{p}{R}{R-1}{R-2}{u+v}=$$
$$=\sum_{p} \w{R}{R+1}{p}{R-1}{u+v} \w{R-1}{R}{R-1}{p}{u} \w{R-1}{p}{R-1}{R-2}{v},\eqno(3.23)$$}Therefore Eq. (3.21) is obeyed. The rest of the equations may be proved similarly using the relations (3.14-3.20).

It is straitforward to show that the BW of our model enjoy the following properties:

{\it Initial condition:} 

$$\w{a}{b}{c}{d}{0}=\delta_{b,d}, \eqno(3.24)$$

{\it Reflection symmetry:}

$$\w{a}{b}{c}{d}{u}=\w{a}{c}{b}{d}{u}, \eqno(3.25)$$

{\it Rotational symmetry: }

$$\w{a}{b}{c}{d}{u} = \sqrt{{G_{b} G_{d}} \over  {G_{a} G_{c}}} \w{d}{a}{b}{c}{-\lambda-u}, \hskip 0.8cm G_{j} \equiv (2-\delta_{j,R}-\delta_{j,R^{\prime}}) [(j+{1 \over 2}) \omega], \eqno(3.26)$$

{\it Two inversion relations:}

$$\sum_{g} \w{a}{g}{c}{d}{u}  \w{a}{b}{c}{g}{-u}  =\delta_{b,d} {{[\lambda+u][\lambda-u][\omega+u][\omega-u]} \over {[\lambda]^2 [\omega]^2}}, \eqno(3.27)$$

$$\sum_{g} \wl{a}{b}{g}{d}{\lambda-u}  \wl{c}{d}{g}{b}{\lambda+u}  =\delta_{b,d}{{[\lambda+u][\lambda-u][\omega+u][\omega-u]} \over {[\lambda]^2 [\omega]^2}}, \eqno(3.28) $$where

$$\wl{a}{b}{c}{d}{u} \equiv \sqrt{{G_{a} G_{c}} \over {G_{b} G_{d}}} \w{a}{b}{c}{d}{u} . \eqno(3.29)$$

\section{Discussion}

We built the solvable lattice model starting  from the rational conformal field theory given by WZW model on $SO(3)_{4 R}=SU(2)_{k=4 R} / Z_{2}$. Note that although the RCFT we started with was defined for $R$ even, the solution of the Yang-Baxter equation so obtained is also valid for  $R$ odd.

 We note that the result obtained here seems to be closely related to that of \cite{ginfendl}, where by the orbifold procedure in the context of the lattice models the direct application of the relations similar to Eqs.(3.14-3.20) was implied. In this way many of the known ADE lattice models \cite{ade} were found to be related in a simple manner \cite{ginfendl}.

The construction applied here \cite{gep} admits natural generalization for the higher representations (for the fused graphs) as well as the generalization to the higher rank, for example for the models based on $SU(n)_{k=n R} / Z_{m}$ ($m$ divisor of $n$) extended current algebra \cite{bg}.

\vskip 1cm

{\hskip 1cm {\bf ACKNOWLEDGMENTS}}

\vskip 1cm

The author thanks Doron Gepner for numerous useful discussions.


\def\mm#1#2#3#4{{\pmatrix{#1 & #2 \cr #3 & #4 }}}
\def\mmm#1#2#3#4#5#6#7#8#9{{\pmatrix{#1 & #2 & #3 \cr #4 & #5 & #6 \cr #7 & #8 & #9}}}

\vskip 1cm

{\hskip 1cm  {\bf APPENDIX}}

\vskip 1cm

Here we list the explicit expressions for the projectors $P^{(j)}$ from \cite{pasq}:

$$P^{(2)}={\bf 1}-P^{(0)}-P^{(1)}, \hskip 0.8cm P^{(i)} P^{(j)} =\delta_{i,j} P^{(j)}. \eqno(A1)$$The matrix elements of $P^{(0)}$ and $P^{(1)}$ ($\langle a,b,d | P^{(j)} | a,c,d \rangle$) are given by:

$P^{(0)}$:

$$\langle a,b,d | P^{(0)} | a,c,d \rangle=0, \hskip 0.8cm a {\not =} d, \eqno(A2)$$

$$\langle j,b,j |P^{(0)}| j,c,j \rangle={1 \over {[3] [2 j+1]}} \mmm{[2 j-1]}{*}{*}{\sqrt{[2 j-1][2 j+1]}}{[2 j+1]}{*}{\sqrt{[2 j-1][2 j+3]}}{\sqrt{[2 j+1][2 j+3]}}{[2 j+3]}, \hskip 0.8cm b,c=j-1,j,j+1 \eqno(A3) $$

$P^{(1)}$:

$$\langle j,b,j+1 |P^{(1)}| j,c,j+1 \rangle={{[2]} \over {[4] [2 j +2]}} \mm{[2 j]}{*}{\sqrt{[2 j] [2 j + 4] }}{[2 j + 4]}, \hskip 0.8cm b,c=j,j+1, \eqno(A4)$$

$$\langle j,b,j-1 |P^{(1)}| j,c,j-1 \rangle={{[2]} \over {[4] [2 j]}} \mm{[2 j+2]}{*}{-\sqrt{[2 j+2] [2 j -2] }}{[2 j -2]}, \hskip 0.8cm b,c=j,j-1, \eqno(A5)$$

{\small $$\langle j,b,j |P^{(1)}| j,c,j \rangle={{[2]} \over [4]} \mmm{(1-{[2] \over {[2 j] [2 j +1]}})}{*}{*}{-\sqrt{{[2 j-1]} \over {[2 j+1]}} {{q^{2 j+1}+q^{-2 j-1}} \over {[2 j]}}}{{2+q^{4 j+2}+q^{-4 j-2}} \over {[2 j][2 j+2]}}{*}{-\sqrt{[2 j+3][2 j-1]} \over [2 j+1]}{\sqrt{{[2 j+3]} \over {[2 j+1]}} {{q^{2 j+1}+q^{-2 j-1}} \over {[2 j+2]}}}{(1-{[2] \over {[2 j+2] [2 j +1]}})}, b,c=j-1,j,j+1. \eqno(A6)$$}


\end{document}